\documentclass{amsart}

\usepackage{amssymb,amsfonts,latexsym,amscd,epsfig}

\begin{document}

\def\A{{\mathcal A}}
\def\alg{{\mathfrak A}}
\def\amp{{\rm Amp}}
\def\ampi{{A_{\dex,\tau,J,\lambda,\e}}}

\def\Bound{{\mathcal B}}

\def\bra{\big\langle}

\def\cj{K_\dex(J)}
\def\C{{\Bbb C}}
\def\Ch{C^{(h)}}
\def\tCh{\tilde C^{(h)}}
\def\Cv{C^{(v)}}

\def\Detau{{\alg}_L(I_\tau)}
\def\cDetau{{\alg}_L(I_\tau^c)}
\def\Dell{{\alg}_{L}^{(\omega)}(\e,\delta,\ell;I_\tau)}
\def\cDell{\Detau\setminus\Dell}
\def\Dom{\mathfrak{Dom}}
\def\dist{{\rm dist}}
\def\dex{{\sigma}}

\def\Exp{{\Bbb E}}
\def\Expnd {{\Bbb E}_{n-d}}
\def\Exptc{{\Bbb E}_{2-conn}}
\def\Expd{{\Bbb E}_{disc}}
\def\e{\varepsilon}
\def\en{{e_\Delta}}

\def\Fou{{\mathcal F}}

\def\H{{\mathcal H}}
\def\HLL{{H_\omega^{(\LL)}}}
\def\Hpl{{\Bbb H}}

\def\ids{{\overline{{\rm IDS}}}}

\def\lb{\left[}
\def\Ie{I}
\def\Im{{ Im}}

\def\Je{J_{\lambda^2}}
\def\Jeta{J_\eta}

\def\ket{\big\rangle}

\def\LL{\Lambda_L}
\def\tLL{\tilde \LL}
\def\LLinv{\frac{1}{|\LL|}}

\def\mat{{\mathcal M}}
\def\mes{{\rm mes}}
\def\mset{K}

\def\N{{\Bbb N}}

\def\p{r}
\def\pip{\tau}

\def\qm{q^{(m+1)}}

\def\rb{\right]}
\def\R{{\Bbb R}}
\def\Rem{ R}

\def\Sc{{\mathcal S}}
\def\supp{{\rm supp}}

\def\tk{{\tilde k}}
\def\Tor{\Bbb T}
\def\tr{{\rm Tr}}

\def\uj{\underline{j}}
\def\up{\underline{\vp}}
\def\uk{\underline{\vk}}
\def\utk{\underline{\tilde\vk}}
\def\uvw{\underline{\vw}}
\def\ux{\underline{x}}

\def\Z{{\Bbb Z}}

\def\vx{{ x}}
\def\vy{{ y}}
\def\ve{{ e}}
\def\vk{{ k}}
\def\vl{{ l}}
\def\vm{{ m}}
\def\vn{{ n}}
\def\vp{{ p}}
\def\vQ{{ Q}}
\def\vq{{ q}}
\def\vr{{ r}}
\def\vv{{ v}}
\def\vw{{ w}}

\def\1{{\bf 1}}

\def\eqnn{\begin{eqnarray*}}
\def\eeqnn{\end{eqnarray*}}
\def\eqn{\begin{eqnarray}}
\def\eeqn{\end{eqnarray}}
\def\bal{\begin{align}}
\def\eal{\end{align}}

\def\prf{\begin{proof}}
\def\endprf{\end{proof}}

%\def\prf{\noindent{\em Proof.}$\;$}
%\def\endprf{\hspace*{\fill}\mbox{$\Box$}}

%%\numberwithin{equation}{section}

\newtheorem{theorem}{Theorem}[section]
\newtheorem{corollary}{Corollary}[section]
\newtheorem{definition}{Definition}[section]
\newtheorem{proposition}{Proposition}[section]
\newtheorem{lemma}{Lemma}[section]

\title{Localization lengths for Schr\"odinger
operators on $\Z^2$ with decaying random potentials}

\author{Thomas Chen}

\address{Department of Mathematics,
Princeton University,
807 Fine Hall, Washington Road,
Princeton, NJ 08544, U.S.A.}

\email{tc@math.princeton.edu}

\date{}

\maketitle

\begin{abstract}
We study a class of Schr\"odinger operators on $\Z^2$
with a random potential decaying as $|x|^{-\dex}$, $0<\dex\leq\frac12$,
in the limit of small disorder strength $\lambda$.
For the critical exponent $\dex=\frac12$, we prove that
the localization length of eigenfunctions is bounded below by
$2^{\lambda^{-\frac14+\eta}}$,
while for $0<\dex<\frac12$, the lower bound is
$\lambda^{-\frac{2-\eta}{1-2\dex}}$, for any $\eta>0$.
These estimates "interpolate" between the lower bound $\lambda^{-2+\eta}$
due to recent work of Schlag-Shubin-Wolff for $\dex=0$,
and pure a.c. spectrum for $\dex>\frac12$ demonstrated in recent work of Bourgain.
\end{abstract}

\parskip = 8 pt

\section{Introduction}

We study the discrete random Schr\"odinger operator
\eqn
        H_\omega=\Delta+\lambda V_\omega
\eeqn
on $\ell^2(\Z^2)$,
where $\Delta$ is the (centered) nearest neighbor Laplacian, with
spectrum $[-4,4]$, and $\lambda$ is a small parameter
(the disorder strength).
The random potential is given by $V_\omega(x)=v_\dex(x)\omega_x$,
where $v_\sigma(x)\sim|x|^{-\dex}$
and $\{\omega_x\}_{x\in\Z^2}$ are Gaussian i.i.d. random variables.
The restriction to Gaussian randomness has expository advantages,
but is not essential for our techniques to apply.
Extension of our methods to non-Gaussian random potentials can be
accessed along the lines demonstrated in \cite{ch}.
The purpose of this paper is to derive lower bounds on the localization
lengths of eigenfunctions of $H_\omega$.

In the supercritical case $\dex>\frac12$,
it was proven by Bourgain in \cite{bo1} that with large probability,
$H_\omega$ (with Bernoulli or Gaussian randomness)
has, for small $\lambda$, pure a.c. spectrum in $(-4+\tau,-\tau)\cup(\tau,4-\tau)$
($\tau>0$ arbitrary, but fixed); moreover,
the wave operators were constructed, and asymptotic completeness was established.
The (generalized) eigenfunctions are therefore delocalized.
Certain other classes of lattice Schr\"odinger operators
with decaying random potentials have been proven to exhibit a.c. spectrum,
scattering, and asymptotic completeness by Bourgain in \cite{bo2},
and by Rodnianski and Schlag in \cite{rosc}. We also note the contextually
related work of Denissov in \cite{de}.

In the case $\dex=0$, Schlag, Shubin and Wolff have proven lower bounds
on the localization length of eigenfunctions of the form
$\lambda^{-2+\eta}$, for any $\eta>0$, \cite{shscwo}.
For $\dex=0$ and $d=3$, lower bounds of the form
$\lambda^{-2}|\log\lambda|^{-1}$ were derived in \cite{ch}.

We shall here address the case $0<\dex\leq\frac12$ in dimension two.
Our main results are as follows.

For the critical decay exponent $\dex=\frac12$, the problem is
{\em marginal} in the language of renormalization group theory.
Accordingly, we obtain a comparison of the {\em logarithm}
of the localization length to powers of $\lambda$,
yielding lower bounds on the localization length
that are {\em exponential} in $\frac1\lambda$, of the form
$2^{\lambda^{-\frac14+\eta}}$ ($\eta>0$ arbitrary).

In the subcritical case $0<\dex<\frac12$,
it is suspected that the model exhibits a significant component
of point spectrum. In the language of renormalization group theory,
the potential scales like a {\em relevant} perturbation,
whereby we obtain a comparison of
the localization length to powers of $\lambda$.
Consequently, our lower bounds on the localization lengths are
{\em polynomial} in $\frac1\lambda$ for $0<\dex<\frac12$,
of the form $\lambda^{-\frac{2-\eta}{1-2\dex}}$ ($\eta>0$ arbitrary).

On the one hand, our strategy employs graph expansion methods
due to Erd\"os and Yau \cite{erd, erdyau}, and further elaborated on
by the author \cite{ch, ch1}. On the other hand, we use a
smoothing of resolvent multipliers by dyadic restriction,
inspired by Bourgain's approach in \cite{bo1}.
Our methods can be extended to higher dimensions,
but we will here only focus on the case $d=2$.

The following works, which determine macroscopic hydrodynamic limits of the quantum
dynamics in the Anderson model at small disorders (without spatial decay, i.e. $\dex=0$),
are closely related to the topics discussed here.
In an important early work, Spohn proved in \cite{sp} that the kinetic macroscopic
scaling and low coupling limit is determined by a linear Boltzmann equation,
locally in macroscopic time. Erd\"os and Yau proved the corresponding global in
macroscopic time result for the continuum model in $\R^d$, $d=2,3$,
and Gaussian randomness, \cite{erdyau}, which was extended by Erd\"os to the case of a
Schr\"odinger electron interacting with a phonon heat bath, \cite{erd}. The author
derived the corresponding result for the lattice $\Z^3$ and non-Gaussian
randomness, \cite{ch}, and proved that the mode of convergence can be
extended to $r$-th mean, for any $r\in\R_+$ (the previous works proved
convergence in expectation), \cite{ch1}.
Eng and Erd\"os proved the corresponding result for the kinetic macroscopic
and low density limit, \cite{engerd}.
Very recently, Erd\"os, Salmhofer and Yau established the breakthrough
result that beyond kinetic scaling, the macroscopic dynamics is governed
by a diffusion equation, \cite{erdsalmyau}.

\section{Definition of the model and statement of the main results}
\label{intro-sect-1}

We consider the discrete random Schr\"odinger operator
\eqn
        H_\omega = \Delta  + \lambda V_\omega \;
        \label{Homega-def}
\eeqn
on $\ell^2(\Z^2)$, with a radially decaying potential function
\eqn
            V_{\omega}(x) =v_\dex(x) \omega_x \;,
\eeqn
where
$\{\omega_x\}_{x\in\Z^2}$ are independent, identically distributed Gaussian random variables
normalized by $\Exp[\omega_x]=0$, $\Exp[\omega_x^2]=1$, for all $x\in\Z^2$.
Expectations of higher powers of $\omega_x$ satisfy Wick's theorem,
see \cite{erdyau}, and our discussion below.

We shall use the convention
\eqn
            \Fou(f)(k)\;\equiv\;\hat f(k)&=&\sum_{x\in\Z^2}
            e^{-2\pi i kx} f(\vx)
            \nonumber\\
                \Fou^{-1}(g)(x)\;\equiv\;\check g(x)
            &=&\int_{\Tor^2} dk\,
            e^{2 \pi i k x} g(k)
\eeqn
for the Fourier transform and its inverse,
where $\Tor:=[-\frac12,\frac12]$.

We introduce a partition of unity $\sum_{j=0}^\infty P_j=1$ on $\Z^2$,
where  $P_j\sim \chi(2^j<|x|\leq2^{j+1})$, $j\in\N_0$,
is an approximate characteristic functions for a dyadic shell of scale $2^j$.
We require that $|\Fou(P_j P_{j'})|$, for $|j-j'|\leq1$,
are bump functions on $\Tor^2$ at the dual scale $2^{-j}$ satisfying
$\|\Fou(P_j P_{j'})\|_{L^1(\Tor^2)}\sim 1$.
We shall assume that $v_\dex$ is such that
for any $j,j'\in\N_0$ with $|j-j'|\leq 1$, the Fourier transform of $P_j P_{j'} v_\dex^2$ satisfies
\eqn
        |\Fou(P_j P_{j'} v_\dex^2)| \leq C
        2^{-2\dex j}|\Fou(P_j P_{j'} )|
        \sim C  2^{-2\dex j}|\Fou(P_j^2)|  \;,
        \label{Fouv-dyad-est-1}
\eeqn
for a constant $C$ independent of $j,j'$.
Since
\eqn
        \|P_jv_\dex\|_{\ell^\infty(\Z^2)}=
        \|P_j^2 v_\dex^2\|_{\ell^\infty(\Z^2)}^{1/2}
        \leq\|\Fou(P_j^2 v_\sigma^2)\|_{L^1(\Tor^2)}^{1/2}
        \sim 2^{-\dex j}\;,
\eeqn
this in particular implies that
\eqn
        |x|^{\dex}|v_\dex(x)|\leq C \;,
\eeqn
for $0<\dex\leq\frac12$.

The centered nearest neighbor lattice Laplacian $\Delta$
defines the Fourier multiplier
\eqn
        \Fou(\Delta  f) (\vk) =   \en(\vk) \hat f(\vk) \;,
\eeqn
where
\eqn
        \en(\vk) = 2\cos(2\pi k_1)+2\cos(2\pi k_2)
        \label{kinendef}
\eeqn
is the quantum mechanical kinetic energy of the electron.

For almost every realization of $V_\omega$, $H_\omega$ is a selfadjoint operator
on $\ell^2(\Z^2)$.

We shall use the same argument for the determination of the localization
length of eigenfunctions of $H_\omega$ as in \cite{ch}.
Let $L> e^{\lambda^{-2}}$, and
\eqn
        \Lambda_L:=[-L,L]^2\cap\Z^2 \;.
\eeqn
For $\ell\ll L$ and $x\in\LL$, let
\eqn
        R_{x,\delta, \ell}\sim \chi\big(\,\big\{y\in\Z^2\big|\,
        \frac{\delta\ell}{2}<|x_i-y_i|<\frac{\ell}{2}\,,\,
        i=1,2\big\}\,\big)
\eeqn
denote an approximate  characteristic function supported on a
cubical shell centered at $x$, of outer and inner side lengths $\ell$ and
$\delta\ell$, respectively. We shall adopt
the choice for $R_{x,\delta,\ell}$ from \cite{ch},
which is a product of differences of Fej\'er kernels with
\eqn
        \|R_{x,\delta,\ell}\|_{\ell^\infty(\LL)}=1 \;.
\eeqn
It is not necessary here to specify $R_{x,\delta,\ell}$ in more detail, as
its explicit form only enters a result that can be
straightforwardly adapted from \cite{ch} (Eq. (~\ref{free-evol-est-1})).

Given a fixed realization of the random potential for which
$H_\omega$ is selfadjoint on $\ell^2(\Z^2)$,
let $\HLL$ denote the restriction of $H_\omega$ to $\LL$.
Moreover, let $\{\psi_\alpha^{(L)}\}_{\alpha\in\alg_L}$ denote an orthonormal $\HLL$-eigenbasis in
$\ell^2(\LL)$
\eqn
        (\HLL\psi_\alpha^{(L)})(x)&=&e_\alpha^{(L)}\psi_\alpha^{(L)}(x)
        \;\;\;
        (x\in\LL) \;,
\eeqn
satisfying Dirichlet boundary conditions
\eqn
        \psi_\alpha^{(L)}(x)=0
        \;\;\;
        (x\in\partial\LL
        :=\Lambda_{L+1}\setminus\LL)\;.
        \label{eigenL}
\eeqn
The number of eigenfuntions is given by
\eqn
        |\alg_L|=|\LL| \;.
\eeqn
Let, for $\tau>0$ arbitrary but fixed, and independent of $\lambda$ and $\dex$,
\eqn
        I_\tau:= (-4+\tau,-\tau)\cup(\tau,4-\tau) \;.
\eeqn
Let
\eqn
        \Detau:=\{\alpha\in\alg_L\big|\,e_\alpha^{(L)}\in
        I_\tau\}\;,
\eeqn
and similarly as in \cite{ch}, let
for $\e$ small
\eqn
        \Dell&:=&\big\{\,\alpha\in\Detau\big|\,
        \nonumber\\
        &&\hspace{1cm}
        \sum_{x\in\LL} |\psi_\alpha^{(L)}(x)| \,
        \big\| R_{x, \delta, \ell} \psi_\alpha^{(L)}
        \big\|_{\ell^2(\LL)} < \e\,\big\}  \;.
\eeqn
As pointed out in \cite{ch}, the key observation is that
$\{\psi_\alpha^{(L)}\}_{\alpha\in\Dell}$
contains the class of localized eigenstates
with energies in $I_\tau$ that are concentrated in
balls of radius $O(\frac{ \delta \ell }{ \log \ell })$,
with $\delta$ independent of $\ell$.

Our main result is the following theorem.

\begin{theorem}
\label{thm-main-1}
For $\delta>0$ sufficiently small, $0<\lambda\ll\delta$,
any fixed $\tau$ with $\lambda\ll\tau<\delta$,
and any arbitrary $\eta>0$,
\eqn
        \liminf_{L\rightarrow\infty}\Exp \left[
        \frac {|\alg_L\setminus\alg_L(\delta^{\frac45},\delta,\ell_\dex(\lambda);I_\tau)|}
        {|\alg_L|} \right]\ge 1 - \delta^{\frac{1}{5}} \;.
    \label{thm-main-est-1}
\eeqn
The lower bound on the localization length $\ell_\dex(\lambda)$
satisfies the following estimates:
\begin{itemize}
\item
In the subcritical case $0<\dex<\frac12$,
there exist  positive constants  $\lambda_0(\dex,\eta)\ll1$ and $C_\dex$
for every fixed $0<\dex<\frac12$ such that
\eqn
    \ell_\dex(\lambda)\geq C_\dex \lambda^{-\frac{2-\eta}{1-2\dex}}
\eeqn
for all $\lambda<\lambda_0(\dex,\eta)$.
\item
In the critical case $\dex=\frac12$, there exists a positive constant
$\lambda_0(\eta)\ll1$ such that
\eqn
    \ell_{\dex=\frac12}(\lambda)\geq   2^{\lambda^{-\frac14+\eta}}
\eeqn
for all $\lambda<\lambda_0(\eta)$.
\end{itemize}
\end{theorem}

\noindent{We} add the following remarks.
\begin{itemize}
\item
(~\ref{thm-main-est-1}) trivially implies
\eqn
    {\Bbb P}\Big[\liminf_{L\rightarrow\infty}
    \frac {|\alg_L\setminus\alg_L(\delta^{\frac45},\delta,\ell_\dex(\lambda);I_\tau)|}
        {|\alg_L|}>1-\delta^{\frac{1}{10}}\Big]>1-\delta^{\frac{1}{10}} \;.
\eeqn

\item
Spectral restriction to the interval $I_\tau$  suppresses infrared singularities, and
enables one to apply certain smoothing procedures to $\frac{1}{\en-z}$,  \cite{bo1}.

\item
Only a slight modification of the bounds used in our analysis of the subcritical case
along the lines of \cite{ch} is necessary to yield
the lower bound $\lambda^{-2+\eta}$ for $\dex=0$. Inclusion of a classification of graphs
argument as in \cite{erdyau, ch} would improve the lower bound to
$\lambda^{-2}|\log\lambda|^{-1}$.
We shall not further discuss these matters here, since the argument is the same as
the one presented in \cite{ch} for the 3-D problem.
\end{itemize}

\section{Proof of Theorem {~\ref{thm-main-1}}}

Our starting point is the following key lemma. It is an extension of
a joint result with L. Erd\"os and H.-T. Yau in \cite{ch}.

\begin{lemma}\label{ceylemma}
Let $\e,\delta>0$ be small and $\lambda\ll1$.
Assume that there exists $t^*(\delta,\ell)>0$, such that
\eqn
    \label{mainest}
        &&\Exp \Big[\frac{1}{|\alg_L|}
        \sum_{x\in\LL}\big\| R_{x, \delta, \ell}\chi_{I_\tau}(\HLL)
        e^{-i t^*(\delta,\ell) \HLL }
        \delta_x\big\|_{\ell^2(\LL)} ^2\Big]
        \nonumber\\
        &&\hspace{3cm}\ge 1- \e - \Exp\Big[\frac{|\cDetau|}{|\alg_L|}\Big]
        -C\frac{\ell}{L}   \;.
\eeqn
Then,
\eqn
        \liminf_{L\rightarrow\infty}\Exp\left[
        \frac {|\alg_L\setminus \Dell| } {|\alg_L|}\right]\ge 1 - 4
        \e^{\frac12}\;.
\eeqn
\end{lemma}

\prf
The proof follows closely a line of arguments presented in \cite{ch},
but comprises key modifications due to the restriction of
the energy range to $I_\tau$.

We expand $\delta_x$ in the eigenbasis $\{\psi_\alpha^{(L)}\}$,
\eqnn
        \delta_x &=& \sum_\alpha {a_x^\alpha} \psi_\alpha^{(L)}
        \; \;\\
        a_x^\alpha &=& \overline{\big\langle \delta_\vx \, , \,
        \psi_\alpha^{(L)} \big\rangle }
        = \overline{\psi_\alpha^{(L)}(x)} \;,
\eeqnn
so that in particular,
\eqn
        \|\delta_x\|_{\ell^2(\LL)}^2=\sum_{\alpha\in\alg_L}|a_x^\alpha|^2=1\;.
        \label{axalphl2norm}
\eeqn
Applying the Schwarz inequality,
\eqn
        \Big\| R_{x, \delta, \ell}\chi_{I_\tau}(\HLL) e^{-i t \HLL }
        \delta_x\Big\|_{\ell^2(\LL)}^2
        \leq (1+  \e^{-\frac12} )(A)+
        (1+\e^{\frac12}) (B) \; ,
\label{CSest1}
\eeqn
where
\eqn
        (A)&:=&\Big\|R_{x, \delta, \ell}e^{-i t \HLL }
        \sum_{\alpha \in \Dell}{a_x^\alpha}
        \psi_\alpha^{(L)} \Big\|_{\ell^2(\LL)}^2
        \nonumber\\
        &\leq& \Big\|R_{x, \delta, \ell}
        \sum_{\alpha \in \Dell}
        e^{-i t e_\alpha^{(L)} }{a_x^\alpha}
        \psi_\alpha^{(L)}\Big\|_{\ell^2(\LL)}
        \nonumber\\
        &\leq&  \sum_{\alpha\in \Dell}
        |\psi_\alpha^{(L)}(x)| \big\|  R_{x, \delta, \ell}
        \psi_\alpha^{(L)} \big\|_{\ell^2(\LL)} \;,
\eeqn
using the a priori bound
\eqn
        (A)&\leq&
        \Big\|
        \sum_{\alpha \in \Dell}e^{-i t e_\alpha^{(L)} }{a_x^\alpha}
        \psi_\alpha^{(L)} \Big\|_{\ell^2(\LL)}^2
        \nonumber\\
        &=&\sum_{\alpha\in\Dell}|a_x^\alpha|^2 \;\leq \; 1
        \;,
\eeqn
which follows from $\|R_{x, \delta, \ell}\|_\infty=1$,
orthonormality of
$\{\psi_\alpha^{(L)}\}_{\alpha\in\alg_L}$,
and (~\ref{axalphl2norm}).

Moreover,
\eqn
        (B)&:=&\Big\| R_{x, \delta, \ell} e^{-i t \HLL }
        \sum_{\alpha \in \cDell}
        {a_x^\alpha} \psi_\alpha^{(L)}
        \Big\|_{\ell^2(\LL)}^2
        \nonumber\\
        &\leq&
        \Big\|\sum_{\alpha \in \cDell}e^{-i t e_\alpha^{(L)} }
        {a_x^\alpha} \psi_\alpha^{(L)} \Big\|_{\ell^2(\LL)}^2
        \nonumber\\
        &=&
        \sum_{\alpha \in \cDell}|a_x^\alpha|^2
        \nonumber\\
        &=&
        \sum_{\alpha \in \cDell}
        |\psi_\alpha^{(L)}(x)|^2 \; .
\eeqn
Summing over $x\in\LL$,
\eqn
        \sum_{x\in\LL} \big\| R_{x, \delta, \ell}
        e^{-i t \HLL }
        \delta_x\big\|_{\ell^2(\LL)}^2
        &\leq& (1+\e^{\frac12})\, \big|\Detau\setminus\Dell\big|
        \label{DbarDsplitest}\\
        &+&\e (1+ \e^{-\frac12} )\, |\Dell|  \; ,
        \nonumber
\eeqn
using the definition of $\Dell$.

Let $I_\tau^c:=\R\setminus I_\tau$. We thus get
\eqn
        \frac {|\alg_L\setminus\Dell | } {|\alg_L|}
        &=&\frac{|\cDetau|}{|\alg_L|}
        + \frac {|\Detau\setminus\Dell | } {|\alg_L|}
        \nonumber\\
        &\geq&  \frac{|\cDetau|}{|\alg_L|}
        \nonumber\\
        &+&
        \frac{1-\e^{\frac12}}{|\alg_L|}\sum_{x\in\LL}  \big\| R_{x, \delta, \ell}
        \chi_{I_\tau}(\HLL)
        e^{-i t \HLL } \delta_x\big\|_{\ell^2(\LL)}^2
        \nonumber\\
        &-& (1+ \e^{-\frac12})\, \e  -C\frac{\ell}{L} \;.
        \label{fracAcAlowbd}
\eeqn
Taking expectations and using  (~\ref{mainest}),
\eqn
        \Exp\Big[\frac {|\alg_L\setminus\Dell | } {|\alg_L|}\Big]
        &\geq&1-\e^{\frac{1}{2}}\Exp\Big[\frac{|\cDetau|}{|\alg_L|}\Big]-3\e^{\frac12}
        -C\frac{\ell}{L}\;.
\eeqn
Since $\frac{|\cDetau|}{|\alg_L|}\leq1$, this implies the claim.
\endprf

Our strategy therefore is to find large values for $\ell$ and $t^*(\delta,\ell)$
such that (~\ref{mainest}) is satisfied.

The following lemma controls the free Schr\"odinger evolution.

\begin{lemma}\label{fundestlemma}
Let for $\lambda$ small and $0<\delta<1$
\eqn
        t^*(\delta ,\lambda):=\delta^{\frac45}\ell \;.
\eeqn
Then, the free evolution satisfies
\eqn\label{fundest0}
        && \frac{1}{|\alg_L|}\sum_{x\in\LL}
        \big\|  R_{x,\delta,  \ell_\dex(\lambda)}\chi_{I_\tau}(\HLL)
        e^{-i t^*(\delta, \lambda) \Delta } \delta_x \big\|_{\ell^2(\LL^2)}^2
        \nonumber\\
        &&\hspace{3.5cm}\geq 1 -  \delta^{\frac{3}{10}} - \frac{|\cDetau|}{|\alg_L|}
        -C\frac{\ell }{L}\;.
\eeqn
\end{lemma}

\prf
We note that
\eqn
    &&\sum_{x\in\LL}
        \big\|  R_{x,\delta,  \ell_\dex(\lambda)}\chi_{I_\tau}(\HLL)
        e^{-i t^*(\delta, \lambda) \Delta } \delta_x \big\|_{\ell^2(\LL^2)}^2
        \nonumber\\
        &&\hspace{2cm}\geq \;(I)-(II)
\eeqn
where
\eqn
    (I)&:=&\sum_{x\in\LL}\|R_{x,\delta,  \ell_\dex(\lambda)}
    e^{-i t^*(\delta, \lambda) \Delta } \delta_x \big\|_{\ell^2(\LL)}^2
    \nonumber\\
    (II)&:=&\sum_{x\in\LL} \big\|\chi_{I_\tau^c}(\HLL)
    e^{-i t^*(\delta, \lambda) \Delta } \delta_x \big\|_{\ell^2(\LL)}^2  \;.
\eeqn
This follows from $\chi R^2\chi=\chi^2-\chi\overline{R^2}\chi=
1-\overline{\chi^2}-\chi\overline{R^2}\chi
\geq 1-\overline{R^2}-\overline{\chi^2}=R^2-\overline{\chi^2}$, where
$R\equiv R_{x,\delta,  \ell_\dex(\lambda)}$, $\chi\equiv \chi_{I_\tau}(\HLL)$,
and $\bar{A}:=1-A$ (so that $\overline{\chi^2}=\chi_{I_\tau^c}^2(\HLL)$).

Replacing $\|\,\cdot\,\|_{\ell^2(\LL)}$ by $\|\,\cdot\,\|_{\ell^2(\Z^2)}$
in $(I)$ costs a boundary term of size $O(\ell L)$ or smaller. Since $|\alg_L|\sim L^2$,
\eqn
    &&\frac{1}{|\alg_L|}\sum_{x\in\LL}\|R_{x,\delta,  \ell }
        e^{-i t^*(\delta,\lambda)\Delta } \delta_x \big\|_{\ell^2(\LL)}
    \nonumber\\
    &=&\frac{1}{|\alg_L|}\sum_{x\in\LL}\|R_{x,\delta,  \ell }
        e^{-i t^*(\delta,\lambda)\Delta } \delta_x \big\|_{\ell^2(\Z^2)}
    +O(\frac{\ell}{L}) \;.
\eeqn
We then find
\eqn
    \|R_{x,\delta,  \ell }
    e^{-i t^*(\delta,\lambda)\Delta } \delta_x \big\|_{\ell^2(\Z^2)}
    \geq 1-\delta^{\frac{3}{10}} \;,
    \label{free-evol-est-1}
\eeqn
from a related argument in \cite{ch}, adapted to the present case.

On the other hand,
\eqn
    (II)&\leq&\sum_{x\in\LL}
    \big\|   \chi_{I_\tau^c}(\HLL)
    e^{-i t^*(\delta, \lambda) \Delta } \delta_x \big\|_{\ell^2(\LL)}^2
    \nonumber\\
    &=&\tr\Big[e^{i t^*(\delta, \lambda) \Delta }\chi_{I_\tau^c}(\HLL)e^{-i t^*(\delta, \lambda) \Delta }\Big]
    \nonumber\\
    &=&\tr\Big[\chi_{I_\tau^c}(\HLL) \Big]
    \nonumber\\
    &=&|\cDetau| \;.
\eeqn
Recalling that $|\LL|=|\alg_L|$, this completes the proof.
\endprf

Our result is implied by the following key lemma. It controls the
interaction of the electron with the impurity potential over
a time $t^*$ comparable to the lower bound on the localization
length $\ell_\dex(\lambda)$.

\begin{lemma}
\label{Lemma-main-0}
Let for $0<\delta<1$
\eqn
    t^*_{\delta,\dex,\lambda}=\delta^{\frac45}\ell_{\dex}(\lambda) \;.
    \label{tstar-def-1}
\eeqn
Then, for any arbitrary, but fixed $\tau>0$,
\eqn\label{fundest10}
        &&\limsup_{L\rightarrow\infty}
    \Exp  \Big[\frac{1}{|\alg_L|}\sum_{x\in\LL}\big\|\chi_{I_\tau}(\HLL)\big(
    e^{-i t^*_{\delta,\dex,\lambda} \HLL }\delta_x-
        e^{-i t^*_{\delta,\dex,\lambda} \Delta } \delta_x
    \big) \big\|_{\ell^2(\LL)}^2\Big]
        \nonumber\\
    &&\hspace{3cm}\leq C\tau^{\frac12}+\lambda^{\eta}
    \, \;.
\eeqn
The definition of $\ell_\dex(\lambda)$ is given in
Theorem {~\ref{thm-main-1}}.
\end{lemma}

To establish Lemma {~\ref{Lemma-main-0}},
it suffices to prove the following estimate.

\begin{lemma}
\label{Lemma-main-1}
Under the assumptions of Lemma {~\ref{Lemma-main-0}},
\eqn
    &&\sup_{\phi\in\ell^2(\Z^2)\atop\|\phi\|_{\ell^2(\LL)}=1}
    \Exp\big[\|\chi_{I_\tau}(H_\omega)\big(
    e^{-i t^*_{\delta,\dex,\lambda} H_\omega }-
    e^{-i t^*_{\delta,\dex,\lambda} \Delta }
    \big)\phi\|_{\ell^2(\Z^3)}^2\big]<C\tau^{\frac12}+\lambda^{\eta}\;.
    \label{Lemma-main-est-1}
\eeqn
\end{lemma}

The rest of this paper is devoted to the proof of Lemma {~\ref{Lemma-main-1}}.

\section{Resolvent expansion }

Let henceforth $t\equiv t^*_{\delta,\dex,\lambda}$.
We write
\eqn
    \phi_t=\chi_{I_\tau}(H_\omega)e^{-itH_\omega}\phi_0
\eeqn
with $\phi_0\in\ell^2(\Z^2)$
in resolvent representation
\eqn
    \phi_t&=&\frac{1}{2\pi i}e^{\e t}\int_{\R} d\alpha e^{-it \alpha}
    \frac{\chi_{I_\tau}(H_\omega)}{H_\omega-\alpha-i\e}\phi_0
\eeqn
where we will use the choice
\eqn
    \e=\frac1t
\eeqn
in all that follows.
Due to the spectral restriction of $H_\omega$ to the disjoint union of
intervals $I_\tau$, the $\alpha$-integration
contour can be deformed into
\eqn
    \phi_t&=&\frac{1}{2\pi i}e^{\e t}\int_{C_-\cup C_+}
    d\alpha e^{-it \alpha}
    \frac{\chi_{I_\tau}(H_\omega)}{H_\omega-\alpha-i\e}\phi_0\;,
\eeqn
where the loops
\eqn
    C_-&:=&[-4+\tau/2,-\tau/2]\cup(-4+\tau/2-2i\e[0,1])\cup
    \nonumber\\
    &&([-4+\tau/2 ,-\tau/2 ]-2i\e)\cup(-\tau/2-2i\e[0,1])
    \nonumber\\
    C_+&:=&[\tau/2,4-\tau/2]\cup( 4-\tau/2-2i\e[0,1])\cup
    \nonumber\\
    &&([\tau/2 ,4-\tau/2 ]-2i\e)\cup(\tau/2-2i\e[0,1])
\eeqn
are taken in the clockwise direction.
$C_-$ and $C_+$ each enclose one of the components of $I_\tau-i\e$.

Let $\Cv:=\{\Cv_j\}_{j=1}^4$ denote the four vertical,
and $\Ch:=\{\Ch_j\}_{j=1}^4$ the four horizontal segments in $C_-$ and $C_+$.
Each segment carries an orientation accounting for the direction in which
the contour integration is taken.

Then,
\eqn
    |\frac{1}{2\pi i}e^{\e t}\int_{\Cv_j} d\alpha e^{-it \alpha}
    \frac{\chi_{I_\tau}(H_\omega)}{H_\omega-\alpha-i\e}\phi_0|
    &<&\frac14 |\Cv_j|\sup_{z\in S_j\atop z'\in I_\tau-\e}|z-z'|
    \nonumber\\
    &=&\e\tau^{-1}\;,
\eeqn
as $\dist(\Cv_j,I_\tau-i\e)=\tau/2$,
and $|\Cv_j|=2\e$.

Henceforth, we shall omit the subscript "$\omega$" in the
random potential $V_\omega\equiv V$.

Defining
\eqn
    \phi^{(h)}_t&:=&\frac{1}{2\pi i}e^{\e t}\int_{\Ch}
    d\alpha e^{-it \alpha}
    \frac{1}{H_\omega-\alpha-i\e}\phi_0\;,
\eeqn
we have
\eqn
    \|\phi_t\|_{\ell^2(\Z^2)}^2&\leq&2\Big(\frac\e\tau\Big)^2+
    2\|\chi_{I_\tau}(H_\omega)\phi^{(h)}_t\|_{\ell^2(\Z^2)}^2 \;.
    \label{phi-ell2-ircut-bound-1}
\eeqn
Next, we expand $\phi_t^{(h)}$ into
\eqn
        \phi_t^{(h)}=\sum_{n=0}^N \phi_{n,t}+R_{N,t} \;,
\eeqn
where the $n$-th term is given by
\eqn
        \phi_{n,t}&:=&\frac{e^{\e t}}{2\pi i}\int_{\Ch} d\alpha e^{-it\alpha}
    \tilde\phi_{n,\e}(\alpha)\;,
\eeqn
with
\eqn
    \tilde\phi_{n,\e}(\alpha)&:=&(-\lambda)^n
    \frac{1}{\Delta-\alpha-i\e}
    \Big(V\frac{1}{\Delta-\alpha-i\e}\Big)^n
        \phi_0 \;.
    \label{phiNt-def-1}
\eeqn
In frequency space,
\eqn
        \Fou( \phi_{n,t} )(k_0)&=&\frac{1}{2\pi i} e^{\e t}\int_{\Ch}  d\alpha
        e^{-it\alpha}\Fou(\tilde\phi_{N,\e}(\alpha))(k_0)
\eeqn
where
\eqn
    \Fou(\tilde\phi_{N,\e}(\alpha))(k_0)&=&(-\lambda)^n\int_{(\Tor^3)^n}dk_1\cdots dk_n
    \frac{1}{\en(k_0)-\alpha-i\e}
        \nonumber\\
        &&\times\,
    \Big[\prod_{j=1}^n\frac{1}{\en(k_j)-\alpha-i\e} \hat V(k_{j}-k_{j-1})\Big]
        \hat \phi_0(k_n) \;,
        \label{hatphint-expans}
\eeqn
and $\Tor=[-\frac12,\frac12]$.
We will refer to the Fourier multiplier $\frac{1}{\en(k)-\alpha-i\e}$
as a {\em particle propagator}.

The remainder term is given by
\eqn
        R_{N,t}=- \lambda e^{\e t}\frac{1}{2\pi i} \int_{\Ch}
    d\alpha e^{-it\alpha} \frac{1}{H_\omega-\alpha-i\e} V\tilde\phi_{N,\e}(\alpha) \;.
    \label{RNt-def-1}
\eeqn
The depth of the expansion $N$ remains to be optimized.

We remark that due to the truncation of the integration contour,
$\phi_{n,t}$ and $R_{N,t}$ cannot be written as time integrals
of the form
\eqn
    \phi_{n,t}&\leftrightarrow& (-i\lambda)^n\int_{\R_+^{n+1}}\delta(t-\sum_{j=0}^n s_j)
    e^{-s_0\Delta}V e^{-s_1\Delta}\cdots \cdots V e^{-is_n \Delta}\phi_0
    \nonumber\\
    R_{N,t}&\leftrightarrow&-i\lambda\int_0^t ds e^{-i(t-s)H_\omega}V\phi_{N,s}
\eeqn
as in the Duhamel expansions used in \cite{ch,erd,erdyau,erdsalmyau}.
While for $\phi_{n,t}$, this is not essential in the present work
(because we admit a polynomial error $O(\lambda^\eta)$, $\eta>0$, in
our bounds), our methods require an expression of the above form for $R_{N,t}$
(because we will apply the time partitioning
trick used in \cite{erdyau} and \cite{ch}).

To this end, we claim that
\eqn
    R_{N,t}&=&R_{N,t}^{(0)}+R_{N,t}^{(1)}
    \label{RNt-def-2}
\eeqn
with
\eqn
    R_{N,t}^{(0)}&:=&e^{-itH_\omega}\frac{-\lambda}{2\pi i}\int_{\Ch}
    d\alpha \frac{1}{H_\omega-\alpha-i\e}V\tilde\phi_{N,\e}(\alpha)
    \\
    R_{N,t}^{(1)}&:=&-i\lambda\int_0^tds e^{-i(t-s)H_\omega}V \phi_{N,s}\;.
\eeqn
To see this, we note that (~\ref{RNt-def-1}) implies
\eqn
    \partial_t R_{N,t} = -iH_\omega R_{N,t} -i\lambda V \phi_{N,t} \;,
\eeqn
which is solved by the variation of constants formula (~\ref{RNt-def-2}).

We note that $\chi_{I_\tau}(H_\omega)R_{N,t}^{(0)}$ would vanish if $\Ch$ were
replaced by a connected $\alpha$-integration contour $C_{conn}$ that encloses
$I_\tau-i\e$. This is because
$C_{conn}$ can be deformed into a contour arbitrarily far away from the spectrum of
$\chi_{I_\tau}(H_\omega)H_\omega-i\e$,
as there is no obstructing phase factor $e^{-it\alpha}$.

Furthermore, due to the truncation of the integration contour to $\Ch$, it is
also necessary to control
\eqn
    &&\|\chi_{I_\tau}(H_\omega)\big(\phi_{0,t}-
        e^{-it\Delta}\phi_0\big)\|_{\ell^2(\Z^2)}^2
        \nonumber\\
        &&\hspace{2cm}\leq \int_{\Tor^2}dp
    \Big|\int_{C\setminus \Ch}d\alpha e^{-it\alpha}
        \frac{1}{\en(p)-\alpha-i\e}\Big|^2 \;,
        \label{free-evol-error-1}
\eeqn
where
\eqn
    \tilde C&:=&[-4-\e,4+\e]\cup(4+\e-2i\e[0,1])\cup
    \nonumber\\
    &&\hspace{2cm}([-4-\e,4+\e]-2i\e)\cup(-4-\e-2i\e[0,1])\;.
\eeqn
We write $\tilde C\setminus \Ch =\tilde C_-\cup \tilde C_0\cup \tilde C_+$, where
$\tilde C_{\pm}:=\{z\in \tilde C\setminus\Ch\big|
\pm\Re(z)>2\}$. $\tilde C_-$ and $\tilde C_+$ are connected arcs,
while $\tilde C_0$ consists of two disjoint,
parallel lines, all of length $O(\tau)$. We claim that
\eqn
    &&\Big|\int_{\tilde C_-\cup \tilde C_0\cup \tilde C_+}d
    \alpha e^{-it\alpha}
        \frac{1}{\en(p)-\alpha-i\e}\Big|
        \nonumber\\
        &&\hspace{2cm}<
        C\Big[\chi(|\en(p)+4|<2\tau)+\chi(|\en(p)+4|<2\tau)
        \nonumber\\
        &&\hspace{3cm} +\chi(|\en(p)|<4\tau) + \frac{\e}{\tau}\Big]\;.
\eeqn
For fixed $p$, the size of
\eqn
    \int_{\tilde C_- }d\alpha e^{-it\alpha}
        \frac{1}{\en(p)-\alpha-i\e}
\eeqn
can be estimated as follows.

If $|\en(p)-4|<2\tau$, we deform
$\tilde C_-$ into a loop that encloses $\en(p)-i\e$, and a disjoint arc
of length $O(\e)$ connecting the endpoints of $\tilde C_-$.
The resolvent at $\en(p)-i\e$, due to the loop, yields a factor $e^{-it(\en(p)-i\e)}$.
The integral over the arc is bounded by its length $O(\e)$, multiplied with the bound
$\frac1\e$ on the resolvent. Both contributions are $O(1)$.

If $|\en(p)+4|>2\tau$, we deform $\tilde C_-$
into a line of length $2\e$ connecting its endpoints, which  has a distance
$\geq\tau$ from $\en(p)$. The modulus of the resolvent is therefore
$\leq O(\frac1\tau)$, and integrating, we get an error bound of order $O(\frac\e\tau)$.

The cases $\tilde C_0$ and $\tilde C_+$ are similar.

Thus,
\eqn
    (~\ref{free-evol-error-1})&<&C\Big[\mes\{|\en(p)+4|<2\tau\}+
    \mes\{|\en(p)|<4\tau\}
    \nonumber\\
    &&\hspace{2cm}+\mes\{|\en(p)-4|<2\tau\}+ \frac\e\tau\Big]
    \nonumber\\
    &<& C\tau^{\frac12}\;,
\eeqn
as $\e$ will be chosen $\ll\tau$ in the end.

The Schwarz inequality thus yields
\eqn
        &&\Exp\Big[\|\chi_{I_\tau}(H_\omega)\big(\phi_t^{(h)}-
        e^{-it\Delta}\phi_0\big)\|_{\ell^2(\Z^2)}^2\Big]
    \nonumber\\
    &&\hspace{3cm}\leq\;C\tau^{\frac12} +
        2\, \Exp\Big[ \big\| \sum_{n=1}^N \phi_{n,t} \big\|_2^2 \Big]
        +2 \,\Exp\Big[ \big\| \chi_{I_\tau}(H_\omega)\Rem_{N,t} \big\|_2^2 \Big]
        \nonumber\\
    &&\hspace{3cm}=\;C\tau^{\frac12} +
        2\sum_{n,n'=1}^N \Exp\Big[  \langle\phi_{n',t},\phi_{n,t}\rangle  \Big]
        +2\, \Exp\Big[ \big\| \chi_{I_\tau}(H_\omega)\Rem_{N,t} \big\|_2^2 \Big] \; .
        \label{exp-phi-ell2-Schwarz-1}
\eeqn
Clearly, if $n+n'\not\in2\N$, $\Exp[\langle\phi_{n',t},\phi_{n,t}\rangle]=0$.

We partition $V$ into dyadic shells,
\eqn
    V=\sum_{j=0}^{J+1} V_j \;,
\eeqn
where
\eqn
    V_j(x)&=&P_j(x) v_\dex(x)\omega_x
\eeqn
for $0\leq j\leq J$. The cutoff functions $P_j$ are defined at the beginning of section
{~\ref{intro-sect-1}}. For $j>J$, we rename $P_j\rightarrow \tilde P_j$, and define
\eqn
    P_{J+1}&:=&\sum_{j=J+1}^\infty \tilde P_j
    \label{tildPi-def-1}
\eeqn
Hence, the functions $V_j$ are supported on dyadic annuli of
radii and thicknesses $\sim 2^j$ centered at the origin, $j=1,\dots,J$, while $V_{J+1}$
is the part of $V$ supported in regions with a distance larger than $2^{J+1}$
from the origin.

Let
\eqn
    R_z:=\frac{1}{\Delta-z}\;.
\eeqn
Then, we have
\eqn
       \Exp\lb \langle\phi_{n',t},\phi_{n,t}\rangle \rb &=&
        \sum_{j_1,\dots,j_{2\bar n}=1}^{J+1} \frac{e^{2\e t} \lambda^{2\bar n}}{(2\pi)^2}
        \int_{\Ch\times \overline \Ch} d\alpha d\beta e^{-it(\alpha-\beta)}
        \nonumber\\
        &&\hspace{0.5cm}\Exp\Big[\langle\phi_0\,,\, R_{\alpha+i\e} V_{j_1}
        R_{\beta-i\e}V_{j_2}
        R_{\beta-i\e}\cdots\cdots
        \nonumber\\
        &&\hspace{1.5cm}\cdots\cdots V_{j_n} R_{\beta-i\e}
        R_{\alpha+i\e} V_{j_{n+2}} \cdots
        \cdots V_{j_{2\bar n}}
        R_{\alpha+i\e} \phi_0\rangle\Big]
        \label{exp-phinn-res-1}
\eeqn
for $1\leq n,n' \leq N$, and $\bar n:=\frac{n+n'}{2}\in\N$.
$\overline \Ch$ denotes the complex conjugate of $\Ch$, and is taken in the
counterclockwise direction by the variable $\beta$.

For $1\leq n,n' \leq N$, and $\bar n:=\frac{n+n'}{2}\in\N$, let
\eqn
        \up&=&(p_0,\dots,p_n,p_{n+1},\dots,p_{2\bar n+1})
\eeqn
and
\eqn
        (\alpha_j,\sigma_j) &=& \left\{\begin{array}{ll}(\alpha,1)&
        0\leq j\leq n\\
        (\beta,-1)&n<j\leq 2n+1 \;. \end{array}\right.
\eeqn
Then, in frequency space representation,
\eqn
        (~\ref{exp-phinn-res-1})
        &=& \sum_{j_1,\dots,j_{2\bar n}=1}^{J+1}
        \frac{e^{2\e t} \lambda^{2\bar n}}{(2\pi)^2}
        \int_{\Ch\times \overline \Ch} d\alpha d\beta e^{-it(\alpha-\beta)}
        \nonumber\\
        &&\hspace{0.5cm}
        \int_{(\Tor^3)^{2\bar n+2}} d\up \,\delta (p_n-p_{n+1})
        \overline{\Fou(\phi_0)(p_0)}\Fou(\phi_0)(p_{2\bar n+1})
        \nonumber\\
        &&\hspace{2cm}
        \prod_{l=0}^{2\bar n+1}\frac{1}{\en(p_l)-\alpha_l-\sigma_l\e}
        \nonumber\\
        &&\hspace{3cm}
        \Exp\Big[\prod_{\stackrel{i=1}{i\neq n+1}}^{2\bar n+1}
        \Fou( V_{j_i})(p_i-p_{i-1})\Big]
        \label{exppotgenexpr}
\eeqn
(noting that $\overline{\Fou( V)(k)}=\Fou( V)(-k)$).

\section{Graph expansion}

We systematize the evaluation of the expectation value of products of
random potentials by use of {\em (Feynman) graphs}, which we represent as follows.

We consider two parallel, horizontal solid lines, which we refer to as {\em particle lines},
joined at a distinguished vertex which accounts for the $L^2$-inner product
(henceforth referred to as the "$L^2$-vertex").

The particle line to the left of the $L^2$ vertex shall contain $n$, and the one
its right shall contain $n'$ vertices,
accounting for copies of the random potential $\hat V$ (henceforth referred to
as "$V$-vertices").

The $n+1$ edges on
the left of the $L^2$-vertex
correspond to the propagators in $\hat\psi_{n,t}$, while the
$n'+1$ edges on the right correspond to those in
$\overline{\hat\psi_{n',t}}$. We shall refer to those edges
as {\em propagator lines}.

The expectation produces a sum over the
products of $\bar n=\frac{n+n'}{2}\in\N$ contractions between
all possible pairs of random potentials.
We insert an edge referred to as a {\em contraction line} between every pair of
mutually contracted random potentials. We then identify the contraction type with the
corresponding graph.

We let $\Pi_{n,n'}$ denote the set of all graphs comprising $n+n'$
$V$-vertices, one $L^2$-vertex, two particle lines, $\bar n$ contraction
lines, and $2\bar n+2$ propagator lines as defined above.

An example is given in Figure 1.

\subsection{Dyadic Wick expansion}

We shall next discuss the expectation of products of dyadically resolved
random potentials in detail.

It is evident that
\eqn
        \Exp[V_j(x) V_{j'}(x')]&=&\delta_{|j-j'|\leq1}P_{j}(x)P_{j'}(x)
        v_\dex^2(x)\delta_{x,x'}
        \nonumber\\
        &\leq&C 2^{-2\dex j}\delta_{x,x'}\;,
\eeqn
and
\eqn
        \Exp[V_{J+1}(x) V_{J+1}(x')]&\leq&C 2^{-2\dex J} \delta_{x,x'}\;.
\eeqn
The expectation of products $\prod_i\omega_{x_i}$ satisfies Wick's
theorem, and the same is true for the expectation of
products $\prod_i V_{j_i}(x_i)$. This can be formulated as follows.

There are $\bar n$ pairing contraction lines joining pairs of
$\hat V_\omega$-vertices in $\pi$.
We enumerate the contraction lines in an arbitrary, but fixed order
by $\{1,\dots,\bar n\}$.

We write $i\sim_{m} i'$ to express that
the $i$-th and the $i'$-th $V$-vertex are connected by the $m$-th
contraction line.

Given
\eqn
        \uj&:=&(j_1,\dots,j_{2\bar n})
        \nonumber\\
        \ux&:=&(x_0,\dots,x_{2\bar n+1})\;,
\eeqn
let
\eqn
        \delta_\pi(\uj,\ux) :=\prod_{m=1}^{\bar n}
        \Big[\delta_{|j_{i}-j_{i'}|\leq1}\delta_{x_i,x_{i'}}\Big]\Big|_{i\sim_m i'}\;.
        \label{deltapi-x-def-1}
\eeqn
Then, in position space,
\eqn
    \Exp\Big[\prod_{i=1}^{2\bar n}V_{j_i}(x_i)\Big]=
    \sum_{\pi\in\Pi_{n,n'}}\delta_\pi(\uj,\ux)
    \prod_{i=1}^{2\bar n}v_\dex(x_i)\;.
\eeqn
On the other hand, we arrive at the frequency space picture as follows.

Let
\eqn
        \up&:=&(p_0,\dots,p_n,p_{n+1},\dots,p_{2\bar n+1}) \;.
\eeqn
If $i\sim_m i'$, contraction of
$\Fou(P_{j_i} V)(p_{i+1}-p_{i})$ with
$\Fou(P_{J_{i'}} V)(p_{i'+1}-p_{i'})$ yields
\eqn
        &&\Exp\Big[ \Fou(P_{j_i} V)(p_{i+1}-p_{i})
        \Fou(P_{j_{i'}} V)(p_{ i'+1}-p_{i'})\Big]
        \nonumber\\
        &&\hspace{1.5cm}
        =\delta_{|j_{i}-j_{i'}|\leq1}
        \Fou(P_{j_i}P_{j_{i'}} v_\dex^2)\delta(p_{i+1}-p_{i}+p_{i'+1}-p_{i'})\;.
\eeqn
We define
\eqn
        &&\delta_\pi(\uj,\up;v_\dex):=
        \nonumber\\
        &&\hspace{1cm}\prod_{m=1}^{\bar n}
        \Big[\delta_{|j_{i}-j_{i'}|\leq1}\Fou(P_{j_i} P_{j_{i'}} v_\dex^2)
        \delta(p_{i+1}-p_{i}+p_{i'+1}-p_{i'})\Big]
        \Big|_{i\sim_m i'}\;.
        \label{deltapi-def-1}
\eeqn
Then,
\eqn
    \Exp\Big[\prod_{i=1\atop i\neq n+1}^{2\bar n+1}\Fou(V_{j_i})(p_i-p_{i-1})\Big]=
    \sum_{\pi\in\Pi_{n,n'}}\delta_\pi(\uj,\up;v_\dex) \;.
    \label{Exp-prod-V-1}
\eeqn
We emphasize that the products (~\ref{deltapi-x-def-1}) and (~\ref{deltapi-def-1}) vanish unless the
scales of the contracted dyadic potentials pairwise coincide (up to overlap errors).
That is, $|j_i-j_{i'}|\leq1$ (where $|j_i-j_{i'}|=1$ accounts for overlap errors)
for every pair $i\sim_m i'$.

Expanding the expectation of the product of random potentials,
\eqn
        \Exp[\langle \phi_{n',t},\phi_{n,t} \rangle ]
        &=&\sum_{\pi\in\Pi_{n,n'}}\amp(\pi)
\eeqn
where
\eqn
        \amp(\pi)&=&\sum_{j_1,\dots,j_{2\bar n}=1}^{J+1}
        \frac{e^{2\e t} \lambda^{2\bar n}}{(2\pi)^2}
        \int_{\Ch\times \overline \Ch} d\alpha d\beta e^{-it(\alpha-\beta)}
        \nonumber\\
        &&\hspace{0.5cm}
        \int_{(\Tor^3)^{2\bar n+2}} d\up \,\delta (p_n-p_{n+1})
        \delta_\pi(\uj;\up;v_\dex)
        \nonumber\\
        &&\hspace{3.5cm}
        \overline{\Fou(\phi_0)(p_0)}\Fou(\phi_0)(p_{2\bar n+1})
        \nonumber\\
        &&\hspace{2cm}
        \prod_{l=0}^{2\bar n+1} \frac{1}{\en(p_l)-\alpha_l-\sigma_l\e}
        \; . \label{exp-sum-graphs-1}
\eeqn
Here, $\delta(p_n-p_{n+1})$ corresponds to the $L^2$-vertex.

\section{Bounds on pairing graphs}

We shall use an analogy of the frequency space $L^1-L^\infty$ estimates on the
resolvents adapted to a spanning tree of $\pi$ from \cite{erdyau,ch}.

\begin{lemma}
\label{R-Lnorm-bounds-lemma-1}
Assume that $\alpha\in \Ch$. Then, for the assumptions (~\ref{Fouv-dyad-est-1}) on $P_j$,
\eqn
        &&\Big\| \,\Big|\frac{1}{\en-\alpha-i\e}\Big|*
        |\Fou( P_j P_{j'} v_\dex^2 )|\,\Big\|_{L^\infty(\Tor^2)}
        \nonumber\\
        &&\hspace{3cm}\leq
        \left\{\begin{array}{ll}C_\tau  2^{j(1-2\dex)} &{\rm if}\; j \leq J\\
        C\sigma^{-1}2^{-2\dex J} \e^{-1}&{\rm if}\;j,j'=J+1\;,
        \end{array}
        \right.
        \label{res-Linfty-bound-1}
\eeqn
where the constant $C_\tau$ only depends on $\tau$. Furthermore,
\eqn
        \Big\| \,\Big|\frac{1}{\en-\alpha-i\e}\Big|*|\Fou( P_j P_{j'} v_\dex^2)| \, \Big\|_{L^1(\Tor^2)}
        \leq C\log\frac1\e\;.
        \label{res-L1-bound-1}
\eeqn
for $0\leq j,j'\leq J+1$.
\end{lemma}

\prf
We recall that by (~\ref{Fouv-dyad-est-1}),
\eqn
        |\Fou(P_j P_{j'} v_\dex^2)(p)|&\leq&
        C 2^{-2 \dex j}|\Fou(P_j P_{j'} )(p)|
        \sim C 2^{-2\dex j}
        |\Fou(P_j^2)(p)|
        \label{Pj-ass-Rnorm-1}
\eeqn
for $|j-j'|\leq1$, and any $j$. It thus suffices to discuss the diagonal term $j=j'$.

For $\alpha\in\Ch$, it is shown in \cite{bo1} that given our assumptions on $P_j$,
convolution with $|\Fou (P_j^2)|$ acts like a smoothing operator on $\frac{1}{\en-\alpha-i\e}$,
on the scale dual to $2^j$, to the effect that
\eqn
        \Big|\frac{1}{\en-\alpha-i\e}\Big|*|\Fou( P_j^2)|
        \leq \frac{C_\tau}{|\en-\alpha|+\e+2^{-j} }\;.
\eeqn
The $L^\infty$-bounds (~\ref{res-Linfty-bound-1}) for $0\leq j\leq J$ then follow immediately.
For $j=J+1$,
\eqn
        \Big|\frac{1}{\en-\alpha-i\e}\Big|*|\Fou( P_{J+1}^2)|
        &\leq&\Big\|\frac{1}{\en-\alpha-i\e}\Big\|_{L^\infty(\Tor^2)}
        \sum_{i=J+1}^\infty\|\Fou(\tilde P_i^2 v_\dex^2)\|_{L^1(\Tor^2)}
        \nonumber\\
        &\leq&C\e^{-1}\sum_{i=J+1}^\infty 2^{-2\dex i}\|\Fou(\tilde P_i^2)\|_{L^1(\Tor^2)}
        \nonumber\\
        &\leq&C\e^{-1}\dex^{-1}2^{-2\dex J} \;,
\eeqn
as $\|\Fou(P_i^2)\|_{L^1(\Tor^2)}\sim1$ ($\tilde P_i$ is defined in (~\ref{tildPi-def-1})).

The $L^1$-bound (~\ref{res-L1-bound-1}) has been proven in \cite{ch}.
\endprf

\begin{lemma}
\label{amppi-nn-bound-lemma-1}
For $1\leq n,n'\leq N$, $\tau>0$ and $\pi\in\Pi_{n,n'}$, there exists a finite constant
$C_\tau$ depending only on $\tau$ such that defining
\eqn
        \ampi:=C_\tau(\cj \lambda^2\log\frac1\e
        +\e^{-1}\dex^{-1} 2^{-2\dex J}\lambda^2\log\frac1\e)
        \label{ampi-def-1}
\eeqn
and
\eqn
        \cj:=\left\{
        \begin{array}{cl}
        J+1&{\rm if}\;\dex=\frac12\\
        \frac{ 2^{(1-2\dex){J+1}} -1 }{ 2^{(1-2\dex)}-1 }&{\rm if}\;0<\dex<\frac12\;,
        \end{array}\right.
\eeqn
one gets
\eqn
        |\amp(\pi)|<(\log\frac1\e)^2(\ampi)^{\bar n} \;.
\eeqn
\end{lemma}

\prf
We choose a spanning tree
$T$ on $\pi$ that contains all contraction lines between the pairs of random potentials,
and $\bar n$ out of all particle lines. In addition, $T$ shall include those particle
lines labeled by the momenta $p_n,p_{2\bar n+1}$,
but not those labeled by $p_0,p_{n+1}$. We then call $T$ {\em admissible}.
Momenta (resolvents) supported on $T$ are referred to as tree momenta (resolvents),
and momenta (resolvents) supported on its complement $T^c$ are called loop momenta (resolvents).
We shall then group together every tree resolvent with one adjacent
contraction line carrying a factor $\Fou(P_{j_i}P_{j_{i'}}v_\dex^2)$, $|j_i-j_{i'}|\leq1$, and estimate
the corresponding convolution integral of the form (~\ref{convol-est-1}) below.
All loop resolvents supported on $T^c$ are estimated in $L^1(\Tor^2)$.

We recall that
\eqn
        \amp(\pi)&=&\sum_{j_0,\dots,j_{2\bar n+1}=0}^{J+1}
        \frac{e^{2\e t} \lambda^{2\bar n}}{(2\pi)^2}
        \int_{\Ch\times \overline \Ch} d\alpha d\beta e^{-it(\alpha-\beta)}
        \nonumber\\
        &&\hspace{0.5cm}
        \int_{(\Tor^3)^{2\bar n+2}} d\up \,\delta (p_n-p_{n+1})
        \delta_{\pi}(\uj;\up;v_\dex)
        \nonumber\\
        &&\hspace{3.5cm}
        \overline{\Fou(\phi_0)(p_0)}\Fou(\phi_0)(p_{2\bar n+1})
        \nonumber\\
        &&\hspace{2cm}
        \prod_{l=0}^{2\bar n+1} \frac{1}{\en(p_l)-\alpha_l-i\sigma_l\e}
        \; . \label{exp-sum-graphs-2}
\eeqn
for $\uj=(j_1,\dots,j_{2\bar n})$.

We integrate out the variable $p_{n+1}$, and apply the coordinate
transformation $p_j\mapsto p_j+p_n$, for all $j=0,\dots,n-1,n+2,\dots,2\bar n+1$.
It is easy to see that thereby, $\delta_{\pi}(\uj;\up;v_\dex)$ becomes
independent of $p_n$ and $p_{n+1}$.
We obtain
\eqn
        \amp(\pi)&=&\sum_{j_0,\dots,j_{2\bar n+1}=0}^{J+1}
        \frac{e^{2\e t} \lambda^{2\bar n}}{(2\pi)^2}
        \int_{\Ch\times \overline \Ch} d\alpha d\beta e^{-it(\alpha-\beta)}
        \nonumber\\
        &&\hspace{0.5cm}
        \int_{(\Tor^3)^{2\bar n}} d\up' \,
        \delta_{\pi}'(\uj;\up';v_\dex)
        \nonumber\\
        &&\hspace{0.5cm}
        \int_{\Tor^2}dp_n\frac{1}{\en(p_n)-\alpha-i\e}\frac{1}{\en(p_n)-\beta+i\e}
        \nonumber\\
        &&\hspace{3.5cm}
        \overline{\Fou(\phi_0)(p_0+p_n)}\Fou(\phi_0)(p_{2\bar n+1}+p_n)
        \nonumber\\
        &&\hspace{2cm}
        \prod_{l=0\atop l\neq n,n+1}^{2\bar n+1} \frac{1}{\en(p_l+p_n)-\alpha_l-i\sigma_l\e}
        \; , \label{exp-sum-graphs-3}
\eeqn
where
\eqn
        \up':=(p_0,\dots,p_{n-1},p_{n+2},\dots,p_{2\bar n+1})
\eeqn
and
\eqn
        \delta_{\pi}'(\uj;\up';v_\dex):=
        \delta_{\pi}(\uj;\up;v_\dex)\Big|_{p_{n+1}, p_n\rightarrow0 }\;.
\eeqn
Clearly,
\eqn
        |\amp(\pi)|&\leq&\sum_{j_0,\dots,j_{2\bar n+1}=0}^{J+1}
        \frac{e^{2\e t} \lambda^{2\bar n}}{(2\pi)^2}
        \Big[\sup_{q,q'\in\Tor^2}\int_{\Ch\times \overline \Ch} |d\alpha|\,|d\beta|
        \nonumber\\
        &&\hspace{0.5cm}
        \int_{\Tor^2}dp_n\frac{1}{|\en(p_n)-\alpha-i\e|}\frac{1}{|\en(p_n)-\beta+i\e|}
        \nonumber\\
        &&\hspace{3.5cm}
        \Big|\overline{\Fou(\phi_0)(p_0+q)}\Fou(\phi_0)(p_{2\bar n+1}+q')\Big|\Big]
        \nonumber\\
        &&\hspace{0.5cm}
        \sup_{\alpha\in\Ch}\sup_{\beta\in\overline{\Ch}}\sup_{p_n\in\Tor^2}
        \Big[\int_{(\Tor^3)^{2\bar n}} d\up' \,
        \delta_{\pi}'(\uj;\up';v_\dex)
        \nonumber\\
        &&\hspace{3.5cm}
        \prod_{l=0\atop l\neq n,n+1}^{2\bar n+1}
        \frac{1}{|\en(p_l+p_n)-\alpha_l-\sigma_l\e|}  \Big]
        \; . \label{exp-sum-graphs-4}
\eeqn
Thus, dividing the resolvents into tree and loop terms and defining
\eqn
        \delta_\pi(\uj):=
        \prod_{m=1}^{\bar n} \delta_{|j_{i}-j_{i'}|\leq1}
        \Big|_{i\sim_m i'}\;,
        \label{deltapi-def-2}
\eeqn
(see also (~\ref{deltapi-def-1})), one gets
\eqn
        |\amp(\pi)|&\leq&\sum_{j_0,\dots,j_{2\bar n+1}=0}^{J+1}
        \frac{e^{2\e t} \lambda^{2\bar n}}{(2\pi)^2}
        \delta_{\pi}(\uj)
        \nonumber\\
        &&\hspace{0.5cm}
        \Big[\sup_{q,q'\in\Tor^2}\int_{\Tor^2}dp_{n}|\phi_0(p_n+q)|\,|\phi_0(p_n+q')|\Big]
        \nonumber\\
        &&\hspace{1cm}\Big[\sup_{p_n\in\Tor^2}\int_{\Ch}|d\alpha|\,\frac{1}{|\en(p_n)-\alpha-i\e|}
        \nonumber\\
        &&\hspace{2.5cm}\int_{\overline{\Ch}}|d\beta|\,
        \frac{1}{|\en(p_{n})-\beta+i\e|}\Big]
        \nonumber\\
        &&
        \sup_{\alpha\in\Ch}\sup_{\beta\in\overline{\Ch}}\sup_{p_n\in\Tor^2}
        \Big\{\;
        \Big[\prod_{ T^c}\Big\|\frac{1}{\en-\alpha_i\pm i\e}\Big\|_{L^1(\Tor^2)}\Big]
        \nonumber\\
        &&\hspace{2cm}
        \Big[\prod_{ T}
        \Big\|\,\Big|\frac{1}{\en-\alpha_i\pm i\e}\Big|*
        \big|\Fou(P_{j_i}P_{j_{i'}} v_\dex^2)\big|_{i\sim i'}\,\Big\|_{L^\infty(\Tor^2)}\Big] \; \Big\} \;,
\eeqn
where $i\sim i'$ implies that the vertices indexed by $i$ and $i'$ are linked by a contraction line.
$\prod_T$ and $\prod_{T^c}$ denote the products over all resolvents supported on
$T$ and $T^c$, respectively.
Assuming (~\ref{Pj-ass-Rnorm-1}), we can bound the off-diagonal terms $|j_i-j_{i'}|=1$
by the diagonal terms $j_i=j_{j'}$,
and due to Lemma {~\ref{R-Lnorm-bounds-lemma-1}}, we have
\eqn
        \sup_{q\in\Tor^2}
        \int_{\Tor^2} dp\Big|\frac{1}{\en(p)-\alpha-i\e}\Big|\,\big|\Fou(P_j^2 v_\dex^2)(p-q)\big|
        \leq C_\tau 2^{(1-2\dex)j}
        \label{convol-est-1}
\eeqn
if $0\leq j\leq J$, and
\eqn
        \sup_{q\in\Tor^2}
        \int_{\Tor^2} dp\Big|\frac{1}{\en(p)-\alpha-i\e}\Big|\,\big|\Fou(P_{J+1}^2 v_\dex^2)(p-q)\big|
        \leq \e^{-1}\dex^{-1}2^{-2\dex J}
        \label{convol-est-2}
\eeqn
if $j=J+1$. Hence,
\eqn
        |\amp(\pi)|&\leq&(C\log\frac1\e)^2\|\phi_0\|^2_{L^2(\Tor^2)}
        \sum_{j_0,\dots,j_{2\bar n+1}=0}^{J+1}
        \delta_{\pi}(\uj)
        (C\log\frac1\e)^{|T^c|}
        \nonumber\\
        &&\hspace{1cm}\prod_{i=1}^{2\bar n}\Big(2^{(1-2\dex) j_i }\chi(j\leq J) +
        \dex^{-1}\e^{-1}2^{-2\dex J} \delta_{j_i,J+1}\Big)^{1/2}  \;,
        \label{exp-sum-graphs-5}
\eeqn
where we have used
\eqn
        \sup_{p\in\Tor^2}\int_{\Ch}|d\alpha|\,\frac{1}{|\en(p)-\alpha-i\e|}<C\log\frac1\e\;.
\eeqn
The power $\frac12$ on the last line in (~\ref{exp-sum-graphs-5})
arises because the product extends over all random potentials,
while $T$ accounts only for the contraction lines (each adjacing to two random potentials).
We note also that $\delta_\pi(\uj)$ forces elements of $\uj$ to be pairwise equal,
up to overlap terms.

Therefore,
\eqn
        |\amp(\pi)|&\leq&(C\log\frac1\e)^{2+|T^c|}
        \Big(\sum_{j=0}^J 2^{(1-2\dex)j}+\sigma^{-1}\e^{-1}2^{-2\dex J}\Big)^{|T|} \;,
\eeqn
where $|T|$ and $|T^c|$ denote the numbers of resolvents supported on $T$ and $T^c$, respectively.
From
\eqn
        \sum_{j=0}^J 2^{(1-2\dex)j}=\left\{
        \begin{array}{ll}
        J+1&{\rm if}\;\sigma=\frac12\\
        \frac{ 2^{(1-2\dex)(J+1)} -1 }{ 2^{(1-2\dex)}-1 }&{\rm if}\;0<\sigma<\frac12
        \end{array}\right.
\eeqn
and $|T|=|T^c|=\bar n$,
the assertion of the lemma follows.
\endprf

\section{Estimating the remainder term}

The remainder term of the resolvent expansion is given by
\eqn
        R_{N,t}=- \lambda e^{\e t}\frac{1}{2\pi i} \int_{\Ch}
    d\alpha e^{-it\alpha} \frac{1}{H_\omega-\alpha-i\e} V\tilde\phi_{N,\e}(\alpha) \;,
    \label{RNt-def-1-2}
\eeqn
as we recall from (~\ref{RNt-def-1}).
The trivial bound
\eqn
    \Exp[\|R_{N,t}\|_{\ell^2(\Z^2)}^2]&\leq&
    C\lambda^2\e^{-2}\Exp[\|V\phi_{N,t}\|_{\ell^2(\Z^2)}^2]
    \nonumber\\
    &\leq&
    N!\lambda^2\e^{-2} (\log\frac1\e)^2 (\ampi)^N
    \label{RNt-triv-bound-lemma-1}
\eeqn
is insufficient in the subcritical case $0<\dex<\frac12$.
We shall instead apply the time partitioning trick used in \cite{erdyau} and \cite{ch}.
In the critical case $\dex=\frac12$, the time partitioning trick is not effective, but
the trivial bound (~\ref{RNt-triv-bound-lemma-1}) suffices.

\subsection{The subcritical case $0<\dex<\frac12$}

We have
\eqn
    R_{N,t}=R_{N,t}^{(0)}+R_{N,t}^{(1)}
\eeqn
with
\eqn
    R_{N,t}^{(0)}&:=&e^{-itH_\omega}\frac{-\lambda}{2\pi i}\int_{\Ch}
    d\alpha \frac{1}{H_\omega-\alpha-i\e}V\tilde\phi_{N,\e}(\alpha)
    \label{RNt-def-2-0}
    \\
    R_{N,t}^{(1)}&:=&-i\lambda\int_0^tds e^{-i(t-s)H_\omega}V \phi_{N,s}\;,
    \label{RNt-def-2-1}
\eeqn
as was shown in (~\ref{RNt-def-2}).

\begin{lemma}
\label{RNt-0-bound-lemma-1}
\eqn
        \Exp[\|R_{N,t}^{(0)}\|_{\ell^2(\Z^2)}^2]&\leq&
        N!\frac{\lambda^2}{\tau^2} (\log\frac1\e)^2   (\ampi)^N \;,
\eeqn
where $\ampi$ is defined in (~\ref{ampi-def-1}).
\end{lemma}

\prf
We can deform the contour $\Ch$ of the $\alpha$-integration in (~\ref{RNt-def-2-0})
into
\eqn
    \tCh&:=&(-4+\tau/2+i[0,1])\cup([-4+\tau/2,-\tau/2]+i)\cup(-\tau/2+i[0,1])\cup
    \nonumber\\
    &&(\tau/2+i[0,1])\cup([4-\tau/2,\tau/2]-i)\cup(4-\tau/2+i[0,1]) \;,
\eeqn
as there is no obstructing phase factor $e^{-it\alpha}$.
One then immediately sees that
\eqn
    \Exp[\|\chi_{I_\tau}(H_\omega)R_{N,t}^{(0)}\|_{\ell^2(\Z^2)}^2]\leq
    \frac{c\lambda^2}{\tau^2}\Exp[\|V\phi_{N,t}\|_{\ell^2(\Z^2)}^2] \;,
\eeqn
since almost surely,
\eqn
    \Big\|\chi_{I_\tau}(H_\omega)
    \frac{1}{H_\omega-\alpha-i\e}\Big\|_{op}<c\tau^{-1}\;,
\eeqn
for any $\alpha\in \tCh$. We note that by the effect of the
infrared regularization, use of unitarity of $e^{it H}$
in estimating (~\ref{RNt-def-2-0}) is {\em not} penalized by
the usual factor $t^2=\e^{-2}$.
\endprf

Using unitarity in bounding the corresponding quantity for $R_{N,t}^{(1)}$,
however, costs a factor $\e^{-2}$, and we shall use the time partitioning trick
of \cite{erdyau} to account for it.

\begin{lemma} For $1\ll\kappa\ll \e^{-1}$, and $0<\dex<\frac12$,
\eqn
        \Exp[\|R_{N,t}^{(1)}\|_{\ell^2(\Z^2)}^2]&\leq&
        (3\kappa N)^2 (\log\frac1\e)^2\sum_{n=N+1}^{4N-1}
        n! (\ampi)^{n}
        \nonumber\\
        &&+ (4N)!
         \frac{1}{\e^2\kappa^{(1-2\dex)N}} (\log\frac1\e)^2
        C^{4N}(\ampi)^{4N}
        \label{RNt-1-est-1}
\eeqn
\end{lemma}

\prf
The asserted estimate is obtained from application of the time partitioning
trick introduced in \cite{erdyau}. The details for the lattice model
are presented in \cite{ch}, and we shall here only sketch the strategy.

We choose $\kappa\in\N$ with
$1\ll\kappa\ll\e^{-1}$, and partition $[0,t]$ into $\kappa$ subintervals
\eqn
        [0,t]=[0,\theta_1]\cup_{j=1}^{\kappa-1}(\theta_j,\theta_{j+1}]
\eeqn
with $\theta_j=\frac{jt}{\kappa}$, $j=1,\dots,\kappa$.
Thereby,
\eqn
        R_{N,t}^{(1)}=-i\lambda\sum_{j=0}^{\kappa-1}e^{-i(t-\theta_{j+1})H_\omega}
        \int_{\theta_j}^{\theta_{j+1}} ds \, e^{-is H_\omega}V \phi_{N,s} \;.
     \label{RemNt-def-1}
\eeqn
Let
\eqn
        \phi_{n,N,\theta}(s)&=&(-i\lambda)^{n-N}
        \int_{\R_+^{n-N+1}} ds_{0}\cdots ds_{n-N}
        \delta(\sum_{j=0}^{n-N}s_j-(s-\theta))
        \nonumber\\
        &&\times\,
        e^{-is_0 \Delta}V\cdots V e^{-is_{n-N}\Delta}
        V \phi_{N,\theta}\;.
\eeqn
That is, the first $N$ out of $n$ collisions happen in the time interval $[0,\theta]$, while
the remaining $n-N$ collisions occur in the time interval $(\theta,s]$.

Expanding $e^{-isH_\omega}$ in (~\ref{RemNt-def-1}) into a Duhamel series with $3N$ terms and
remainder, we find
\eqn
        R_{N,t}^{(1)}=\tilde R_{N,t}^{(<4N)}+\tilde R_{N,t}^{(4N)}\;,
\eeqn
where
\eqn
        \tilde R_{N,t}^{(<4N)}&=&\sum_{n=N+1}^{4N-1}\tilde\phi_{n,N,t} \;,
        \\
        \tilde\phi_{n,N,t}&:=&-i\lambda
        \sum_{j=1}^{\kappa}
        e^{-i(t-\theta_j)H_\omega}V\phi_{n,N,\theta_{j-1}}(\theta_{j})
\eeqn
and
\eqn
        \tilde R_{N,t}^{(4N)}=-i\lambda \sum_{j=1}^{\kappa}e^{-i(t-\theta_j)H_\omega}
        \int_{\theta_{j-1}}^{\theta_j}ds \;
        e^{-i(\theta_j-s)H_\omega}
        V \phi_{4N,N,\theta_{j-1}}(s) \;.
\eeqn
By the Schwarz inequality,
\eqn
        \|\tilde R_{N,t}^{(<4N)}\|_{\ell^2(\Z^2)} \leq (3N\kappa) \sup_{N<n<4N,1\leq j\leq\kappa}
        \|\lambda V
        \phi_{n,N,\theta_{j-1}}(\theta_{j})\|_{\ell^2(\Z^2)}
        \label{RNt-4N-est-1}
\eeqn
and
\eqn
        \|\tilde R_{N,t}^{(4N)}\|_{\ell^2(\Z^2)} \leq t \sup_{1\leq j\leq\kappa}
        \sup_{s\in[\theta_{j-1},\theta_j]}
        \|\lambda V
        \phi_{4N,N,\theta_{j-1}}(s)\|_{\ell^2(\Z^2)} \;.
        \label{RNt-4N-est-2}
\eeqn
The functions $\phi_{n,N,\theta_{j-1}}(\theta_{j})$ and
$\phi_{4N,N,\theta_{j-1}}(s)$ have the following properties.

The expected value of $|(~\ref{RNt-4N-est-1})|^2$ is bounded by the first term after the
inequality sign in (~\ref{RNt-1-est-1}). This is a straightforward consequence of
Lemma {~\ref{amppi-nn-bound-lemma-1}}. For the detailed argument, see \cite{ch, erdyau}.

It remains to estimate (~\ref{RNt-4N-est-2}).
With $\theta'-\theta=\frac t\kappa$, we find
\eqn
        (\hat\phi_{n,N,\theta}(\theta'))(k_0)
        &=&\frac{i(-\lambda)^{n-N} e^{\frac{\e t}{\kappa}}}{2\pi}
        \int_{\Ie}d\alpha e^{-\frac{i\alpha t}{\kappa}}
        \int_{(\Tor^2)^{n-N+1}} dk_{1}\cdots dk_{n-N}
        \nonumber\\
        &&\times\,
        \frac{1}{\en(k_0)-\alpha-i\kappa\e}\hat V(k_1-k_0)\cdots
        \nonumber\\
        &&\hspace{1.5cm}\cdots\,
        \frac{1}{\en(k_{n-N})-\alpha-i\kappa\e}
        \hat V(k_{n-N+1}-k_{n-N})
        \nonumber\\
        &&\times\, \hat\phi_{N,\theta}(k_{n-N+1}) \;,
\eeqn
where we recall that
\eqn
        \hat\phi_{N,\theta}(k_{n-N+1})&=&
        \frac{i(-\lambda)^N e^{\e\theta}}{2\pi}\int_{\Ch}d\alpha e^{-i\theta \alpha}
        \int_{(\Tor^2)^N}\prod_{j=n-N+1}^{n+1}dk_j
        \nonumber\\
        &&\times\,\frac{1}{\en(k_{n-N+1})-\alpha-\frac i\theta}
        \hat V(k_{n-N+2}-k_{n-N+1})\cdots
        \nonumber\\
        &&\hspace{1cm}\cdots\, \hat V(k_{n+1}-k_{n})
        \frac{1}{\en(k_{n+1})-\alpha- \frac i\theta}
        \hat\phi_{0}(k_{n+1}) \;.
\eeqn
The key observation here is that there are $n-N+1$ propagators with
imaginary parts $\pm i\kappa\e$ in the denominator, where $\kappa\e\gg\e$
(and $N+1$ propagators
whose denominators have an imaginary part $-\frac i\theta$, where $\frac1\theta$
and $\e$ can have a comparable size). For those $n-N+1$ propagators,
we have a bound
\eqn
        \frac{1}{|\en-\alpha-i\kappa\e|}*|\Fou(P_j^2 v_\dex^2)|\leq
        C 2^{-2\dex j}\frac{1}{|\en(p)-\alpha|+\kappa\e+2^{-j}}   \;.
\eeqn
We now separate the dyadic scales of the random potential into
\eqn
        0\leq j\leq J'+1 \; \; , \; \; 2^{J'}\sim \frac{1}{\kappa}2^J \;.
\eeqn
Using
\eqn
        \Big\|\,\frac{1}{|\en-\alpha-i\kappa\e|}*|\Fou(P_j^2 v_\dex^2)|
        \,\Big\|_{L^\infty(\Tor^2)} \leq 2^{(1-2\dex)j}
\eeqn
for $j\leq J'$, we have
\eqn
        \sum_{j=0}^{J'}\Big\|\,\frac{1}{|\en-\alpha-i\kappa\e|}*|\Fou(P_j^2 v_\dex^2)|
        \,\Big\|_{L^\infty(\Tor^2)}&\leq& \frac{2^{(1-2\dex)(J'+1)}-1}{2^{(1-2\dex)}-1}
        \nonumber\\
        &\sim& \frac{1}{\kappa^{1-2\dex}}\frac{2^{(1-2\dex)(J+1)}-1}{2^{(1-2\dex)}-1}
        \;.
\eeqn
Furthermore,
\eqn
        \sum_{j=J'+1}^{J+1}\Big\|\,\frac{1}{|\en-\alpha-i\kappa\e|}*|\Fou(P_{j}^2 v_\dex^2)|
        \,\Big\|_{L^\infty(\Tor^2)} &\leq& \frac{1}{\kappa \e}
        \dex^{-1}2^{-2\dex J'}
        \nonumber\\
        &\sim&\frac{1}{\kappa^{1-2\dex}}
        \dex^{-1}\e^{-1}2^{-2\dex J}
\eeqn
for $j=J'+1$.

Therefore, the estimates for resolvents with $\pm i\kappa\e$ in the denominators are
by a factor $\frac{1}{\kappa^{(1-2\dex)}}$ smaller than those for resolvents with $\pm i\e$
derived above.
\eqn
        &&\sum_{j=0}^{J'+1}\Big\|\,\frac{1}{|\en(p)-\alpha-i\kappa\e|}*|\Fou(P_j^2 v_\dex^2)|
        \,\Big\|_{L^\infty(\Tor^2)}
        \nonumber\\
        &&\hspace{3cm}\leq \frac{1}{\kappa^{1-2\dex}}
        \Big(\cj+\sigma^{-1}2^{(1-2\dex)J}\Big)\;.
        \label{treeres-kappa-est-1}
\eeqn
As before, we systematize the evaluation of
\eqn
    \Exp\Big[\|\lambda V
        \phi_{4N,N,\theta_{j-1}}(s)\|_{\ell^2(\Z^2)}^2\Big]
\eeqn
by invoking a graph expansion with $\pi\in\Pi_{4N,4N}$.

For every graph, we again introduce an admissible spanning
tree $T$, as in the proof of Lemma {~\ref{amppi-nn-bound-lemma-1}},
and use the estimate (~\ref{treeres-kappa-est-1})
for tree propagators with $\pm i\kappa\e$ in the denominators.
By the pigeonhole principle, there are at least $N$ of those for every $\pi$, and
any admissible spanning tree $T$ for $\pi$.
This gains a factor of at least
$\frac{1}{\kappa^{(1-2\dex)N}}$ in comparison to the bound in Lemma
{~\ref{amppi-nn-bound-lemma-1}}.
The $L^1(\Tor^2)$-bounds
on loop resolvents are estimated by $C\log\frac1\e$, as before.
Observing that the number of tree propagators is $\bar n$, and that there
are $\bar n+2$ propagators estimated in $L^1$, one concludes that the expected value of
$|(~\ref{RNt-4N-est-2})|^2$ is bounded by the second term after the inequality sign in
(~\ref{RNt-1-est-1}). A detailed exposition is given in
\cite{erdyau} and \cite{ch}.
\endprf

\subsection{The critical case $\dex=\frac12$}

The time partitioning only provides a logarithmic
improvement in $\kappa$,
\eqn
        \sum_{j=0}^{J'}\Big\|\,\frac{1}{|\en-\alpha-i\kappa\e|}*|\Fou(P_j^2 v_\dex^2)|
        \,\Big\|_{L^\infty(\Tor^2)}&\leq& J'+1
        \;\sim\; \frac{1}{\log\kappa}J
\eeqn
which is too small to produce a significant effect.
However, the trivial estimate (~\ref{RNt-triv-bound-lemma-1}) is sufficient
for our analysis,
because the large factor $2^J$ enters $\ampi$ only logarithmically.

\section{Conclusion of the proof of Lemma {~\ref{Lemma-main-1}}}

To conclude the proof of Lemma  {~\ref{Lemma-main-1}}, we make the following choices for
$\e,J,N,\kappa$ as functions of $\dex$, $\lambda$ and $\eta$ (depending implicitly on $\tau$).

\subsection{The subcritical case $0<\dex<\frac12$}

Recalling (~\ref{phi-ell2-ircut-bound-1}), (~\ref{exp-phi-ell2-Schwarz-1}), and
summarizing the estimates formulated in Lemmata {~\ref{amppi-nn-bound-lemma-1}} and
{~\ref{RNt-0-bound-lemma-1}},
our analysis infers that
\eqn
        l.h.s.\;of\;(~\ref{Lemma-main-est-1})
        &<&C\tau^{\frac12}+2\Big(\frac{\e}{\tau}\Big)^2+
        \sum_{ n=1}^N n! (\log\frac1\e)^2 (\ampi)^{ n}
        \nonumber\\
        &&+N!\frac{\lambda^2}{\tau^2}(\log\frac1\e)^2
        (\ampi)^{N}
        \nonumber\\
        &&+\lambda^2 (3\kappa N)^2(\log\frac1\e)^2\sum_{n=N+1}^{4N-1}
        n! (\ampi)^{n}
        \nonumber\\
        &&+(4N)!\frac{\lambda^2}{\e^2\kappa^{(1-2\dex)N}}
        (\log\frac1\e)^2
         (\ampi)^{4N} \;,
\eeqn
where we recall from (~\ref{ampi-def-1}) that
\eqn
        \ampi=C_\tau(\cj \lambda^2\log\frac1\e
        +\e^{-1}\dex^{-1}2^{-2\dex J}\lambda^2\log\frac1\e) \;.
\eeqn
We have
\eqn
        K_{\dex}(J)=\frac{ 2^{(1-2\dex)(J+1)} -1}{2^{1-2\dex}-1}\;.
\eeqn
Let $\eta>0$ be arbitrary but fixed. Setting
\eqn
        \e&=&2^{-J}
        \label{eps-def-lambda-1}\\
        JK_{\dex}(J)&=&\lambda^{-2+2\eta}
        \label{eps-def-lambda-2}
\eeqn
we find
\eqn
        K_{\dex}(J)\lambda^2\log\frac1\e&=&J K_\dex(J)\lambda^2\;\leq\;\lambda^{2\eta}
        \nonumber\\
        \e^{-1}\dex^{-1}2^{-2\dex J}\lambda^2\log\frac1\e&=&
        \dex^{-1}2^{(1-2\dex)J}\lambda^2\log\frac1\e
        \nonumber\\
        &=&\dex^{-1}J K_\dex(J) \;,
\eeqn
so that
\eqn
        \ampi&<&\lambda^{ 1.9\eta } \;,
\eeqn
for $\lambda$ sufficiently small (depending on $\dex$).

Choosing
\eqn
        N&=&\frac{\eta\log\frac1\lambda}{10\log\log\frac1\lambda}
        \;,
\eeqn
one gets (noting that $\e>\lambda^2$)
\eqn
        \sum_{ n=1}^N n! (\log\frac1\e)^2 (\ampi)^{ n}
        &<&C(\log\frac1\lambda)^2\sum_{ n=1}^N  (N\ampi)^{ n}
        \nonumber\\
        &<&C(\log\frac1\lambda)^2\sum_{ n=1}^N  \lambda^{1.5 \eta n}
        \;<\;\lambda^{1.1\eta}
\eeqn
and
\eqn
        N!\frac{\lambda^2}{\tau^2}(\log\frac1\e)^2
        (\ampi)^{N}&<&C(\log\frac1\lambda)^2
        (N\ampi)^N
        \;<\;\lambda
\eeqn
for $\tau\gg\lambda$.
Choosing
\eqn
        \kappa&=&(\log\frac1\lambda)^{\frac{30}{\eta(1-2\dex)}} \;,
\eeqn
one gets
\eqn
        \lambda^2 (3\kappa N)^2(\log\frac1\e)^2\sum_{n=N+1}^{4N-1}
        n! (\ampi)^{n}&<&C\lambda^2 (\log\frac1\lambda)^{\frac{100}{(1-2\dex)\eta}}
        (4N\ampi)^N
        \nonumber\\
        &<&\lambda^{2\eta} \;.
\eeqn
Furthermore, since
\eqn
        \kappa^{(1-2\dex)N}\;>\;\lambda^{-3}
        \;,
\eeqn
one finds
\eqn
        (4N)!\frac{\lambda^2}{\e^2\kappa^{(1-2\dex)N}}
        (\log\frac1\e)^2
        (\ampi)^{4N}&<&(4N\ampi)^{4N}
        \;<\;\lambda^{2\eta}\;.
\eeqn
Thus, for $\lambda$ sufficiently small (depending on $\dex$ and $\eta$),
\eqn
        l.h.s.\;of\;(~\ref{Lemma-main-est-1})
        <C\tau^{\frac12}+\lambda^{\eta}\;.
\eeqn
Moreover, (~\ref{tstar-def-1}),  (~\ref{eps-def-lambda-1}) and
(~\ref{eps-def-lambda-2}) combined imply
that for every fixed $0<\dex<\frac12$, there exists a positive constant $C_\dex$
such that
\eqn
        \ell_\dex(\lambda)\geq C_\dex\lambda^{-\frac{2-\eta}{1-2\dex}} \;.
\eeqn
This proves the assertion of Lemma {~\ref{Lemma-main-1}} for $0<\dex<\frac12$.

\subsection{The critical case $\dex=\frac12$}

Using (~\ref{phi-ell2-ircut-bound-1}),
(~\ref{exp-phi-ell2-Schwarz-1}), Lemma
{~\ref{amppi-nn-bound-lemma-1}} and
({~\ref{RNt-triv-bound-lemma-1}}),
\eqn
        l.h.s.\;of\;(~\ref{Lemma-main-est-1})
        &<&C\tau^{\frac12}+2\Big(\frac{\e}{\tau}\Big)^2+
        \sum_{ n=1}^N n!  (\log\frac1\e)^2  (\ampi)^{ n}
        \nonumber\\
        &&+N!\frac{\lambda^2}{\tau^2}(\log\frac1\e)^2
        (\ampi)^{N}
        \nonumber\\
        &&+N! \lambda^2 \e^{-2} (\log\frac1\e)^2 (\ampi)^{N} \;.
\eeqn
We have
\eqn
        K_{\frac12}(J)=J+1\;.
\eeqn
Let $\eta>0$ be arbitrary (small)  but fixed. Setting
\eqn
        J&=&N\;=\;\lambda^{-\frac14+\eta}
        \nonumber\\
        \e&=&2^{-\lambda^{-\frac14+\eta}} \;=\;2^{-N}\;=\;2^{-J} \;,
        \label{eps-def-lambda-3}
\eeqn
we get, for sufficiently small $\lambda>0$,
\eqn
        \ampi
        &=&C_\tau\Big(J\lambda^2\log\frac1\e+2\e^{-1}2^{-J}\log\frac1\e\Big)
        \nonumber\\
        &<&2C_\tau N^2\lambda^2
\eeqn
and
\eqn
        N^2\ampi&<&\lambda^{3\eta}\;.
\eeqn
Then,
\eqn
        \sum_{n=1}^N n!(\log\frac1\e)^2(\ampi)^n&<&N^2\ampi+\sum_{n=2}^N N^2(N\ampi)^n
        \nonumber\\
        &<&\lambda^{2\eta}
\eeqn
and
\eqn
        N!\frac{\lambda^2}{\tau^2}(\log\frac1\e)^2(\ampi)^N
        &<&\frac{\lambda^{2}}{\tau^2}N^2(N\ampi)^N
        \;<\;\lambda \;.
\eeqn
Furthermore,
\eqn
        N!\e^{-2}\lambda^2(\log\frac1\e)^2(\ampi)^N&<&
        \lambda^2 N^2  (4N\ampi)^N
        \nonumber\\
        &<&\lambda(4\lambda^{2\eta})^{\lambda^{-\frac14+\eta}}
        \;<\;\lambda \;.
\eeqn
In conclusion,
\eqn
        l.h.s.\;of\;(~\ref{Lemma-main-est-1})
        <C\tau^{\frac12}+\lambda^{\eta}\;.
\eeqn
From (~\ref{tstar-def-1}) and  (~\ref{eps-def-lambda-3}), we infer that
\eqn
        \ell_\dex(\lambda)\geq 2^{-\lambda^{-\frac14+\eta}} \;.
\eeqn
This concludes our proof of Lemma {~\ref{Lemma-main-1}} for $\dex=\frac12$.

\subsection*{Acknowledgements}

I am deeply grateful to H.-T. Yau and L. Erd\"os for their support and
generosity. I have benefitted immensely from numerous discussions with H.-T. Yau
about topics closely related to those studied here
while being at the Courant Institute, NYU, as a Courant Instructor.
I also wish to thank M. Aizenman, S. Denissov, V. Jacsic, and S. Warzel for discussions.
This work was supported by NSF grant DMS-0524909.

\parskip = 0 pt
\parindent = 0 pt

%\end{document}

\newpage

\centerline{\epsffile{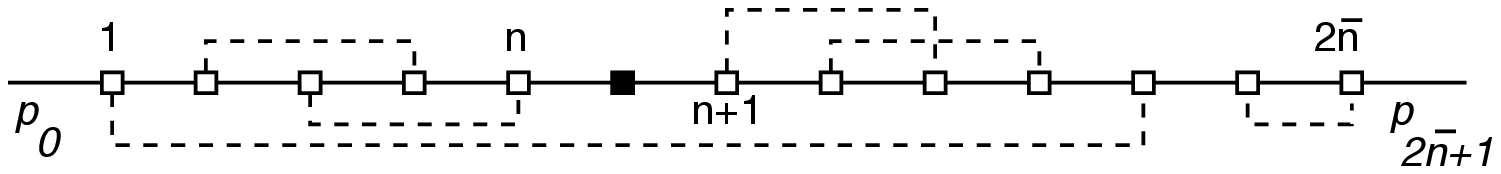} }

\noindent{Figure 1.} A contraction graph $\pi\in\Pi_{n,n'}$ with $n=5,n'=7$ and $\bar n=6$.
The particle lines are solid, the contraction lines dashed. The $L^2$-vertex is black,
while the $V$-vertices are not filled.

\end{document}